\documentstyle[12pt]{article}
\def\be{\begin{equation}}
\def\ee{\end{equation}}
\def\beqn{\begin{eqnarray}}
\def\eeqn{\end{eqnarray}}

\begin{document}
\begin{titlepage}
\title{The Itzykson-Zuber integral for  $U(m|n)$}
\author{Jorge Alfaro \thanks{e-mail address:
jalfaro@lascar.puc.cl .} \ and Ricardo Medina \\
Facultad de F\'\i sica\\Universidad Cat\'olica de
Chile\\Casilla 306, Santiago 22, Chile
\\
and\\
Luis F. Urrutia \thanks{On sabbatical leave from Instituto de Ciencias
Nucleares, UNAM, Mexico.} \\
Facultad de F\'\i sica\\Universidad Cat\'olica de
Chile\\Casilla 306, Santiago 22, Chile\\
and\\
Departamento de F\'\i sica\\
Universidad Aut\'onoma Metropolitana-I\\
Apartado Postal 55-534\\
09340 M\'exico, D.F.}
\date{October 20 1994}
\maketitle
\begin{abstract}
We compute the Itzykson-Zuber(IZ) integral for the superunitary group
$U(m|n)$.  As a consequence, we are able to find the non-zero correlations
of superunitary matrices.

\end{abstract}
\vfill
\begin{flushleft}
PUC-FIS 14\\
October 1994
\end{flushleft}
\end{titlepage}
\newpage

\baselineskip=20 pt
\section{Introduction}

In recent times there has been an enormous amount of work devoted to the
understanding
of random surfaces and statistical systems on random surfaces. The range
of application of these ideas include non-critical string theory as well as
Quantum Chromodynamics (QCD) in the large $N$ limit.
Progress in this area has been possible because the mathematical
knowledge
on random matrices has increased dramatically in the last fifteen years
\cite{mehta1}.

An important mathematical object that appears naturally in the discussion of
random matrices is the integral over the unitary group \cite{IZ}. This integral
has been applied to the solution of the Two matrix model \cite{IZ},
 \cite{mehta2}
and, more recently, to the Migdal-Kazakov model of "induced QCD"
\cite{migkaz}.

On the other hand, there is considerable expectation that supersymmetry
might play an important role in the physical world  \cite{susy}.
Since there is a natural extension of matrices to supermatrices \cite{dewitt},
it is important to understand the properties of random supermatrices and
study their relevance for the theory of random surfaces.

In this paper we will start the study of random supermatrices related to
a supersymmetric extension of the Itzykson-Zuber integral.  In particular we
will calculate some   non-vanishing correlations  of superunitary matrices.
The paper is organized as follows : section 2 contains a very brief review of
some basic properties of Grassmaniann manifolds, which will be used in the
sequel. Section 3 contains the main result of this paper, which is the
calculation of the IZ integral for $U(m|n)$. This result is subsequently used
to find some particular correlations among the  supermatrices elements.

\section{Basic properties of a superspace}

Since the purpose of this paper is to extend the Itzykson-Zuber integral
to the case of the supergroup $U(m|n)$, which can be described by matrices
acting on a superspace (supermatrices), we briefly review some of the
basic properties of the linear algebra together with the differential
and integral calculus defined over a  Grassmann algebra. This sets the stage
for the next section and also fixes our notation. For a more detailed
and complete reference on these matters the reader is referred  to Ref.\cite
{dewitt}.

Let us consider a superspace with coordinates $z^P=(q^i, \theta^\alpha),
i=1,\dots m, \break  \alpha=1, \dots, n$ such that $q^i \ (\theta^\alpha)$
are even \
(odd) elements of a Grassmann algebra. This means that $z^P z^Q=
(-1)^{\epsilon(P) + \epsilon(Q)} z^Q z^P$, where  $\epsilon(P)$ is
the Grassmann
parity of the index $P$ defined by $\epsilon(i)=0,\ mod(2);
\epsilon(\alpha)=
1,\  mod(2)$. Also we have that $\epsilon(z^{P_1}z^{P_2} \dots z^{P_k})=
\sum \epsilon(P_i)$. The above multiplication rule implies in particular
that any odd element of the Grassmann algebra has zero square, i.e. it is
nilpotent.

Supermatrices are arrays that act linearly on the supercoordinates leaving
invariant the partition among  even and odd coordinates. To be more
specific,
the supercoordinates can be thought as forming an
$(m+n)\times 1$ column vector with the first $m$ entries\ (last $ n$
entries)being
even \ (odd) elements of the Grassman algebra. In this way, an $(m+n)\times
(m+n)$ supermatrix is an array written in the partitioned block form

\begin{equation}
M= \left( \begin{array}{ll}
             A  &  B \\
             C  &  D
            \end{array}
\right),
\end{equation}

where the constituent matrices have the entries $A_{ij},\ B_{i\alpha},\
C_{\alpha i}, D_{\alpha\beta}$. Besides, $A_{ij},\ D_{\alpha\beta} \
(B_{i\alpha}, \ C_{\alpha i})$ are even \ (odd) elements of the Grasmann
algebra in  such a way that the  parity array of the supercoordinate vector
columnn is preserved. The parity of any supermatrix element is
$\epsilon(M_{PQ})=
\epsilon(P)+\epsilon(Q)$ and defines the multiplication  among supermatrix
elements. The addition and multiplication of supermatrices
according to the
rules
\begin{eqnarray*}
(M_1+M_2)_{PQ}= (M_1)_{PQ}+(M_2)_{PQ}, \ \ (M_1M_2)_{PQ}=
(M_1)_{PR}(M_2)_{RQ},
\end{eqnarray*}
is such that  it produces again a  supermatrix.
The inverse of a supermatrix  can be constructed in block form, in complete
analogy with the classical  case and it is well defined provided $A^{-1}$
and $D^{-1}$ exist. The inverse of an even matrix is calculated in the
standard way.

The basic invariant of a supermatrix under similarity transformations is the
supertrace
\begin{eqnarray*}
Str(M)=Tr(A)-Tr(D)=\sum_{P=1}^{m+n} (-1)^{\epsilon(P)}M_{PP},\label{str}
\end{eqnarray*}
which is defined so that the cyclic property $Str(M_1M_2)=Str(M_2M_1)$ is
fullfilled for arbitrary supermatrices $M_1, M_2$. The above definition of
the supertrace leads to the construction of the superdeterminant in the form
$Sdet(M)=exp[ Str(ln M)]$, which is explicitly given in the following two
equivalent forms \cite{alemanes}
\begin{eqnarray}
Sdet(M)= \frac {det(A-B D^{-1} C)}{ det(D)}=
\frac {det(A)}{det(D-C A^{-1} B)}.
\end{eqnarray}
The above expression is written only in terms of even matrices in such a way
that the determinant has its usual meaning. The superdeterminat has the
multiplicative property $Sdet(M_1M_2)= Sdet(M_1) Sdet(M_2)$.

The definition of the adjoint supermatrix follows the usual steps by
requiring the identity $(y^{P*} M_{PQ} z^Q)^*=z^{P*}{M^\dagger}_{PQ} y^Q$,
for an arbitrary bilinear form in the complex supercoordinates $y^P$,
 where $*$
denotes complex conjugation. Since the usual definition of complex
conjugation
in a Grassmann algebra $(y^P y^Q)^*=y^{Q*} y^{P*}$ reverses the order of
the factors without introduccing any sign factor , we have the result
${M^\dagger}_{PQ}= {M_{QP}}^*$ as in the standard case.

A hermitian $(m+n)\times(m+n)$ supermatrix $M$ is such that $M^\dagger=M$
and it has $(m+n)^2$ real independent components. The following properties
are also fullfilled:\ (i)\ $(M^\dagger)^\dagger=M$, \ (ii) \ $(M_1M_2)^\dagger=
{M_2}^\dagger {M_1}^\dagger$ and (iii) \  $Sdet(M^\dagger)= {Sdet(M)}^*$.
A unitary $(m+n)\times(m+n)$ supermatrix $U$ is such that $UU^\dagger=
U^\dagger U=I$ ( where $I$ is the identity supermatrix) and also has
$(m+n)^2$ real independent components, which have  the additional
property that
$(SdetU)(SdetU)^*=1$. The set of all  $(m+n)\times(m+n)$ unitary supermatrices
form a group, called the supergroup $U(m|n)$, under the operation of
supermatrix multiplication.

In the next section we will deal with derivation and integration with
respect to the elements of  a supermatrix, which will be considered as
supercoordinates there. To this end,  we summarize here the basic properties
involved, in terms of the supercoordinates $z^A$. Differentiation and
integration over the even supercoordinates follow the same rules and
properties as the corresponding  operations with complex numbers. On the
other hand, the situation with respect to the odd supercoordinates is
greatly simplified because of the nilpotency property,  which leads to the
conclusion that an arbitrary superfunction $F(q^i, \theta^\alpha)$
can be expanded into a finite set of products of the odd supercoordinates.
In fact we have
\begin{eqnarray}
F(q^i, \theta^\alpha)=f(q^i)+ f_{\alpha_1}(q^i) \theta^{\alpha_1} +
f_{{\alpha_1}{\alpha_2}}(q^i)\theta^{\alpha_1}\theta^{\alpha_2} + \dots +
f_{{\alpha_1} \dots {\alpha_n}}(q^i)\theta^{\alpha_1}
\dots \theta^{\alpha_n},
\label{expansion}
\end{eqnarray}
where all the functions $f_{{\alpha_1}\dots {\alpha_k}}$ are completely
antisymmetric in all the subindices. Also, a particular odd supercoordinate
$\theta^{\alpha_k}$ can only appear linearly in the above expansion. Right and
left derivatives with respect to the odd coordinates are defined according
to
\begin{eqnarray}
dF(q^i, \theta_\alpha)={\partial^R F\over \partial \theta^\alpha}
d\theta^\alpha = d\theta^\alpha {\partial^L F \over \partial \theta^\alpha},
\label{partial}
\end{eqnarray}
where the operator $d$
has zero Grassman parity and acts distributively upon a product
of odd coordinates:
 $d(\theta^{\alpha_1}\theta^{\alpha_2}\dots \theta^{\alpha_k})=
{d\theta}^{\alpha_1}\theta^{\alpha_2}\dots \theta^{\alpha_k}+
\theta^{\alpha_1}{d\theta}^{\alpha_2}\dots \theta^{\alpha_k}+
\theta^{\alpha_1}\theta^{\alpha_2}\dots {d\theta}^{\alpha_k}$.
Both, right and left derivatives satisfy adequate Leibnitz rules which can be
directly obtained from the definitions (\ref{partial}). Clearly, for even
coordinates
right and left derivatives coincide.

Integration over odd Grassmann variables is defined, according to Berezin, by
the basic properties \cite {berezin}
\begin{eqnarray}
\int d{\theta^\alpha} =0, \ \ \ \int d \theta^{\alpha} \theta^{\beta}=
\delta^{\alpha\beta},
\end{eqnarray}
which allows the calculation of integrals over arbitrary functions using the
expansion (\ref{expansion}) together with the multiplication rules of
supernumbers. The full
integration measure of the superspace under consideration is given by
$[dz]=dq^1 \dots dq^m d{\theta^1} \dots d{\theta^n}$, where a particular
order in  the odd differentials has been chosen. Under a change of
supercoordinates $z'^{P}=z'^{P}(z^Q)$ the integration measure transforms as
\begin{eqnarray}
[dz']= Sdet( \frac {\partial^R z'^{P} }{\partial z^Q})[dz].
\end{eqnarray}

\section{Integration over the unitary supergroup $U(m|n)$}

In this section we calculate the extension to $(m+n) \times (m+n)$
supermatrices of the IZ integral. Our method, mutatis mutandis, follows
closely to that of Itzykson and Zuber \cite{IZ}. Let us consider the integral
\begin{eqnarray}
\tilde{I} (M_1,M_2;\beta) \equiv \int [dU] e^{\beta Str(M_1 U M_2 U^\dagger)} ,
\label{superint IZ}
\end{eqnarray}
where $M_1, M_2$ are hermitian supermatrices which can be diagonalized
\cite{JAP} and $\beta$ is an even parameter.
Up to a normalization factor $\mu$ (to be fixed later), we define the
integration
measure over $U(m|n)$ by
\begin{eqnarray}
[dU] = \mu \prod_{P,Q=1}^{m+n} dU_{PQ} dU_{PQ}^* \ \delta (UU^\dagger - I).
\label{measure2}
\end{eqnarray}
Here the $\delta$-function really means the product of $(m+n)^2$
unidimensional $\delta$-functions corresponding to the independent
constraints set by the
condition $UU^\dagger = I$. In the case of an odd Grassmann variable
$\theta$, $$\delta (\theta - \bar{\theta}) = (\theta - \bar{\theta}) = \int d
\pi e^{\pi (\theta - \bar{\theta}) },$$
where $\pi$ is another odd integration variable. It is important to observe
that the measure in (\ref{superint IZ}) possesses $2mn$ real independent
odd differentials.

The measure (\ref{measure2}) is invariant under independent right and left
multiplication by arbitrary unitary supermatrices.
In this way, $\tilde{I}$ depends only upon the eigenvalues of $M_1$ and
 $M_2$,
which are given in the corresponding diagonal supermatrices $\Lambda_1$
 and
$\Lambda_2$. Our notation is such that the first $m$ eigenvalues of $\Lambda$
are identified by $\lambda_i$, while the remaining $n$ eigenvalues are
denoted
by ${\bar\lambda}_{\alpha}$. Such partition is characterized by the following
parity assignment of the eigenvector components $V_P,\bar V_P:
\epsilon(V_P)=
\epsilon(P),\epsilon(\bar V_P)=\epsilon(P)+1$.
Before sketching the calculation of the basic integral given in Eq.
 (\ref{superint IZ}), we state our final result and make some general comments.
 We obtain
\begin{eqnarray}
\tilde{I} (\Lambda_1,\Lambda_2;\beta) = \Sigma (\lambda_1,{\bar\lambda}_1)
 \Sigma (\lambda_2,{\bar\lambda}_2) \ \beta^{mn} \times (\beta)^
{- \frac {m(m-1)}{2}} (- \beta)^{- \frac{n(n-1)}{2}}  \times \nonumber \\
\times \prod_{p=1}^{m-1} p! \prod_{q=1}^{n-1} q! \frac {det(e^{ \beta
\lambda_{1i} \lambda_{2j} })}{\Delta (\lambda_1) \Delta (\lambda_2)}
\frac {det(e^{-\beta {\bar\lambda}_{1\alpha} {\bar\lambda}_{2\beta}} )}
{\Delta ({\bar\lambda}_1) \Delta ({\bar\lambda}_2)}.
\label{superint IZ 2}
\end{eqnarray}
Here, $\Delta$ is the usual Vandermonde determinant
\begin{eqnarray}
\Delta (\lambda) = \prod_{i>j} (\lambda_i - \lambda_j) ,\  \
\Delta (\bar\lambda) = \prod_{ \alpha> \beta}
({\bar\lambda}_{\alpha} - {\bar\lambda}_{\beta})
\label{Delta}
\end{eqnarray}
\noindent
and the new function that appears is
\begin{eqnarray}
\Sigma (\lambda,\bar\lambda) = \prod_{i=1}^m \prod_{\alpha=1}^n ( \lambda_i -
{\bar\lambda}_{\alpha} ).
\label{Sigma}
\end{eqnarray}
We observe that the polynomial $\Sigma (\lambda,\bar\lambda)$ is completely
symmetric under independent permutations of $\lambda,\bar\lambda$.

The expression (\ref{superint IZ 2}) is completely determined up to a
normalization factor related to that of the measure in equation
(\ref{measure2}). This situation is analogous to the standard
IZ case where the required factor can be fixed directly from the
corresponding expression by taking the limit
$\Lambda_1, \Lambda_2 \rightarrow 0$ in a convenient way and
demanding $\int [dU] = 1$, for example. This procedure leads to the correct
factors in Eq.(3.4) of Ref.\cite{IZ}.
In our case, a similar limiting procedure leads to the conclusion that $\int
[dU]
\equiv 0,$ precisely due to the appearance of the $\Sigma (\lambda,
\bar{\lambda})$ functions in the numerator. This is not an unexpected
result since we are dealing with odd Grassmann numbers. For this reason
 we have chosen the normalization factor in such a way that
$$\tilde{I} ( \Lambda_1,\Lambda_2;\beta) = \Sigma (\lambda_1,
{\bar\lambda}_1) \Sigma (\lambda_2,{\bar\lambda}_2) \beta^{mn}
I(\lambda_1,\lambda_2;\beta) I({\bar\lambda}_1,{\bar\lambda}_2;-\beta) ,
$$
where $I(d_1,d_2;\beta)$ is the corresponding IZ integral.

Now we give some details of the proof of our result (\ref{superint IZ 2}).
We begin by constructing the superunitary invariant Laplacian operator
 on hermitian supermatrices, which is given by
\begin{eqnarray}
\tilde{D} \equiv \sum_{P,Q=1}^{m+n} (-1)^{\epsilon(P)} \frac {\partial^2}
{\partial M_{PQ} \partial M_{QP}} ,
\label{laplacian superherm}
\end{eqnarray}
where the derivatives are both either left or right derivatives. This is a
particular case of the superlaplacian constructed for curvilinear coordinates
and corresponds to the  metric $ds^2 = Str(dM^2)$ which is invariant under
global unitary transformations $dM \rightarrow U dM U^\dagger$ for hermitian
 supermatrices $M.$

Next we consider the propagator
\begin{eqnarray}
\tilde{f} (M_1,M_2;t) \equiv <M_1| e^{\frac{it}{2} {\tilde{D}}_1} |M_2>,
\label{propagator}
\end{eqnarray}
which, by construction, satisfies the differential equation
\begin{eqnarray}
\left( \frac {\partial}{\partial t} - \frac {i}{2} {\tilde{D}}_1 \right)
\tilde{f}
 (M_1,M_2;t) = 0,
\label{differential equation}
\end{eqnarray}
together with the boundary condition
\begin{eqnarray}
\tilde{f} (M_1,M_2;0) = <M_1|M_2> = \delta (M_1 - M_2),
\label{initial}
\end{eqnarray}
at t=0. It can be shown that the propagator $\tilde{f}$ is given by
\begin{eqnarray}
\tilde{f} (M_1,M_2;t) = \frac {i^{n^2}}{ (2 \pi)^{mn}} \frac {1}{ (2 \pi it)^
{(m-n)^2/2} } e^{\frac {i}{2t} Str(M_1-M_2)^2},
\label{propagator 2}
\end{eqnarray}
in analogy with the standard case.
Since the supertrace of a squared supermatrix is not positive definite,
the  factor $i$ of the exponential in (\ref{propagator}) is included to
insure convergence in further manipulations.

Let us now consider the time evolution of a wave function $\tilde{g}
(M_1;t)$ satisfying (\ref{differential equation}) and such that at $t=0$
reduces to a known function $\tilde{g} (M_1;0) = \tilde{g} (M_1)$, which is
invariant under $U(m|n)$ transformations. As a consequence, $\tilde{g}$ is a
function of  the eigenvalues $\lambda$'s of M only and it is symmetric under
 the separate permutations of the sets $\{ \lambda_i \}$ and
$\{ {\bar\lambda}_{\alpha} \}$. Such a wave function can be constructed
using the propagator in Eq.(\ref{propagator 2}), as
\begin{eqnarray}
\tilde{g} (M_1;t) = \int [dM] \tilde{f} (M_1,M;t) \tilde{g} (M;0),
\label{g(M_1;t)}
\end{eqnarray}
where the  integration measure over hermitian supermatrices is
given by $[dM] = \prod_P dM_{PP} \prod_{S,R>S} dM_{RS} dM_{RS}^*$
and  it is
invariant under   $U(m|n)$ transformations.

Now, we make a change of integration variables, from the initial $M$'s
to   radial ($\Lambda$) and angular ($U$) variables given by
$M = U \Lambda U^\dagger$. This change of variables leads to
\begin{eqnarray}
\tilde{g} (M_1;t) = \int [d\Lambda] [dU] J \tilde{f} (M_1,U
\Lambda U^\dagger;t) \tilde{g} (\Lambda;0) ,
\label{g(M_1;t) 2}
\end{eqnarray}
where the jacobian $J$ is to be determined. The jacobian  factorizes in
the form $J = \tilde{\Delta}^2(\Lambda) J_1(U)$. The piece $J_1(U)$ will
remain incorporated to $[dU]$ while we will find $\tilde{\Delta}^2$
explicitly.
Let us observe that according to Ref.\cite{JAP} there are some restrictions
 to diagonalize a hermitian supermatrix.
We are assuming that such forbidden points constitute a set of zero
measure in the configuration space defined by $\{ M_{PQ} \}$.
In virtue of the invariance of $\tilde{f}$ under $U(m|n)$, with the particular
choice arising from $M_1 = V \Lambda_1 V^\dagger,$ together with the
invariance of $[dU]$ under right and left multiplications, we  arrive at the
result
\begin{eqnarray}
\tilde{g} (\Lambda_1;t) = \int [d\Lambda] [dU] \tilde{\Delta}^2(\Lambda)
 \tilde{f} (\Lambda_1,U \Lambda U^\dagger;t) \tilde{g} (\Lambda),
\label{g(Lambda_1;t)}
\end{eqnarray}
which shows that the wave function preserves its invariance under
 $U(m|n)$ transformations.

The next step is to construct the wave function
\begin{eqnarray}
\tilde{\xi} (\Lambda;t) \equiv \tilde{\Delta} (\Lambda) \tilde{g} (\Lambda;t)
\label{xi}
\end{eqnarray}
and to realize that
\begin{eqnarray}
\tilde{K} (\Lambda_1,\Lambda;t) = \tilde{\Delta} (\Lambda_1) \tilde{\Delta}
 (\Lambda) \int  [dU] \tilde{f} (\Lambda_1,U \Lambda U^\dagger;t),
\label{K}
\end{eqnarray}
is the kernel that corresponds to the propagator of
$\tilde{\xi} (\Lambda;t)$.

Before calculating such kernel in an independent way, so that
we can use Eq.(\ref{K}) to find the value for the angular
integration, we now proceed to the calculation of the jacobian
$\tilde{\Delta}^2.$ This calculation is equivalent to  finding
the volume element in spherical coordinates. The general
property to be used here originates in  the differential geometry
of supermanifolds \cite{dewitt} and  states that given a metric
tensor  such that the length invariant is  $ds^2 = dz^P g_{PQ}
dz^Q,$ the correct integration measure over the manifold is given by
$\sqrt{g} \ [dz]$,  where $g = Sdet(g_{PQ}).$ As we mentioned
previously, the adequate length element in the case of
hermitian supermatrices is $ds^2 = Str(dM^2).$ After making
the change of variables $M = U \Lambda U^\dagger$ we are left with
\begin{eqnarray}
ds^2 = Str(d\Lambda^2) + Str( [U^+ dU, \Lambda]^2 ).
\label{ds^2}
\end{eqnarray}
Since we are only  interested in extracting the $\Lambda$
dependent piece of $\sqrt{g} = \sqrt{g_{\Lambda} g_{U}}$ we
can further consider the representation $U = e^{iH},$ in terms
of another hermitian supermatrix H. Taking H to be infinitesimal
in such a way that $U^\dagger dU = i dH.$, we obtain
\begin{eqnarray}
ds^2 & =  &  \sum_{i,j=1}^m \delta_{ij} d \lambda_i d \lambda_j -
\sum_{\alpha , \beta =1}^n
\delta_{\alpha \beta} d {\bar\lambda}_{\alpha} d {\bar\lambda}_{\beta} +
 \nonumber \\
&   & + \sum_{a,b=1}^m (\lambda_a - \lambda_b)^2 dH_{ab} dH_{ba} +
 \sum_{\alpha ,\beta=1}^n ({\bar\lambda}_{\alpha} - {\bar\lambda}_{\beta})^2
dH_{\alpha \beta} dH_{\beta \alpha} - \nonumber \\
&   & - \sum_{a,\beta =1}^{n,m} (\lambda_a - {\bar\lambda}_{\beta})^2
dH_{a \beta} dH_{\beta a} - \sum_{\alpha ,b =1}^{m,n} ({\bar\lambda}_{\alpha}
- \lambda_b)^2 dH_{\alpha b} dH_{b \alpha},
\end{eqnarray}
for the expression (\ref{ds^2}). From here we conclude that
\begin{eqnarray}
\tilde{\Delta} (\Lambda) = \frac { \prod_{i > j} (\lambda_i - \lambda_j)
\prod_{\alpha > \beta} ({\bar\lambda}_{\alpha} - {\bar\lambda}_{\beta}) }
{ \prod_{i, \alpha } (\lambda_i - {\bar\lambda}_{\alpha}) }={ \Delta(\lambda)
\Delta(\bar \lambda) \over \Sigma(\lambda, \bar \lambda) }.
\label{super Vandermonde}
\end{eqnarray}

Having completely determined the wave function
$\tilde{\xi} (\Lambda;t)$ of eq.(\ref{xi}),
we now
calculate its equation of motion. The starting point is
eq.(\ref{differential equation}) for the function
$\tilde{g} (\Lambda;t) = \tilde{\xi} / \tilde{\Delta} .$
Also we will need the expression of the superlaplacian in
curvilinear coordinates, which is given by
\begin{eqnarray}
\tilde{D} \Psi = \frac {1}{\sqrt{g}} \sum_{P,Q=1}^{m+n} (-1)^{\epsilon_P}
\frac {\partial^L}{\partial z^P} (g^{PQ} \sqrt{g} \frac {\partial^L \Psi}
{\partial z^Q}),
\label{superlaplacian}
\end{eqnarray}
where $g^{PQ}$ is the inverse of the metric tensor $g_{PQ}$.
The expression (\ref{superlaplacian})  further simplifies  in
our case, since we are applying the superlaplacian to a function
which does not depend on the angular variables. In this way we
obtain
\begin{eqnarray}
\frac {\partial \tilde{\xi} (\Lambda;t)}{\partial t} =
\frac {i}{2 \tilde{\Delta} } \left [ \sum_{i=1}^m
\frac {\partial}{\partial \lambda_i} ( \tilde{\Delta}^2
\frac {\partial }{\partial \lambda_i}
(\frac {\tilde{\xi}}{\tilde{\Delta}}) )
- \sum_{i=1}^n \frac {\partial}
{\partial {\bar\lambda}_{\alpha}} ( \tilde{\Delta}^2
\frac {\partial }{\partial {\bar\lambda}_{\alpha}}
(\frac {\tilde{\xi}}{\tilde{\Delta}} ) ) \right],
\label{differential equation 2}
\end{eqnarray}
for the time evolution of  $\tilde{\xi} (\Lambda;t)$, where we have
dropped the refference to left derivatives since all
variables are even. The above equation reduces to
\begin{equation}
\frac {\partial \tilde{\xi} (\Lambda;t)}{\partial t} =
\frac {i}{2} \left[ \sum_{i=1}^m
\frac {\partial ^2}{\partial {\lambda_i}^2} -
\sum_{\alpha=1}^n
\frac {\partial ^2}{\partial {\bar\lambda}_{\alpha}^2} \right] \tilde{\xi}
(\Lambda;t),
\label{differential equation 3}
\end{equation}
in virtue of the following property of the function $\tilde{\Delta}$
\begin{equation}
\left ( \sum_{i=1}^m \frac {\partial ^2}{\partial {\lambda_i}^2} -
\sum_{\alpha= 1}^n \frac
{\partial ^2 }{\partial {\bar\lambda}_{\alpha}^2} \right ) \tilde{\Delta}
\equiv 0.
\label{differential equation Supervandermonde}
\end{equation}
The above equation can be proved  from the expression
\begin{eqnarray}
\sum_{i=1}^m \frac {\partial^2 \tilde{\Delta}}{\partial {\lambda_i}^2} =
\tilde{\Delta}  [ -2 \sum_{\alpha,i,j \neq i} \frac {1}{(\lambda_i -
{\bar\lambda}_{\alpha}) (\lambda_i - \lambda_j)} + 2 \sum_{i, \alpha} \frac
{1}{(\lambda_i-{\bar\lambda}_{\alpha})^2} + \nonumber \\
+ \sum_{i, \alpha , \beta \neq \alpha} \frac {1}{(\lambda_i-
{\bar\lambda}_{\alpha}) (\lambda_i - {\bar\lambda}_{\beta})}  ] ,
\label{second derivative}
\end{eqnarray}
together with the analogous relation
\begin{eqnarray}
\sum_{\alpha=1}^n \frac {\partial^2 \tilde{\Delta} }{\partial {\bar\lambda}_
{\alpha}^2} = \tilde{\Delta}  [ 2 \sum_{\alpha,i} \frac {1}{(\lambda_i-
{\bar\lambda}_{\alpha})^2} + \sum_{\alpha ,i,j \neq i} \frac {1}{(\lambda_i -
{\bar\lambda}_{\alpha})(\lambda_j - {\bar\lambda}_{\alpha})} + \nonumber \\
+ 2 \sum_{\alpha , \beta \neq \alpha , i} \frac {1}{(\lambda_i -
{\bar\lambda}_{\alpha}) ({\bar\lambda}_{\alpha} - {\bar\lambda}_{\beta})} ].
\label{second derivative 2}
\end{eqnarray}
The calculation of the above equations (\ref{second derivative})
and (\ref{second derivative 2}) has already used the identity
$$ \frac {1}{(\mu_1 - \mu_2)(\mu_1 - \mu_3)} +
\frac {1}{(\mu_2 - \mu_3)(\mu_1 - \mu_2)} + \frac {1}{(\mu_3 -
\mu_1)(\mu_3 - \mu_2)} = 0 ,$$ for the cases
$\mu = \lambda, \bar\lambda .$ The difference in  the LHS
of Eq.(\ref{differential equation Supervandermonde}) further  reduces
to
\begin{eqnarray}
\left ( \sum_{i=1}^m \frac {\partial ^2}{\partial {\lambda_i}^2}
- \sum_{\alpha= 1}^n \frac {\partial^2}{\partial
{\bar\lambda}_{\alpha}^2 } \right) \tilde{\Delta} =
\tilde{\Delta}  \left [\sum_i \sum_{\alpha ,
\beta \neq \alpha} A_{\alpha \beta i} -  \sum_{\alpha }
\sum_{i, j \neq i} B_{ij\alpha} \right ] \label{final},
\end{eqnarray}
with
\begin{eqnarray}
A_{\alpha \beta i} =  \frac { {\bar\lambda}_{\alpha} + {\bar\lambda}_{\beta}
- 2 \lambda_i }{ (\lambda_i-{\bar\lambda}_{\alpha}) (\lambda_i-{\bar\lambda}
_{\beta}) ({\bar\lambda}_{\alpha} - {\bar\lambda}_{\beta}) },
\label{A}
\end{eqnarray}
\begin{eqnarray}
B_{ij\alpha} = \frac {\lambda_i + \lambda_j - 2 {\bar\lambda}_{\alpha} }{
(\lambda_i-{\bar\lambda}_{\alpha}) (\lambda_i-\lambda_j)(\lambda_j -
{\bar\lambda}_{\alpha}) } .
\end{eqnarray}
The  result  given in  Eq.(\ref{differential equation Supervandermonde})
is finally  obtained due to the antisymmetry properties $A_{\alpha\beta i}=
-A_{\beta\alpha i}, B_{ij\alpha}=-B_{ji\alpha}$.

The last step in the proof of Eq.(\ref{superint IZ 2}) is
the observation
that the kernel $K(\Lambda_1,\Lambda_2;t)$ in eq.(\ref{K})
satisfies the following differential
equation
\begin{equation}
\frac {\partial K(\Lambda_1,\Lambda_2;t)}{\partial t} =
\frac {i}{2} \left ( \sum_{i=1}^m
\frac {\partial ^2}{\partial \lambda_{1i}^2} -\sum_{\alpha=1}^n
\frac {\partial ^2}{\partial {\bar\lambda}_{1\alpha}^2}
\right ) K(\Lambda_1,\Lambda_2;t),
\label{super heat kernel 4}
\end{equation}
with the initial condition
\begin{equation}
K(\Lambda_1,\Lambda_2;0)= \delta (\Lambda_1-\Lambda_2).
\end{equation}
Moreover we require that $K(\Lambda_1,\Lambda_2;t)$ be separately
 antisymmetric
under permutations of the $\{ \lambda_i \}$ and the $\{ {\bar\lambda}_
{\alpha} \}$.
Then, the  solution of Eq. (\ref{super heat kernel 4}) is
\begin{equation}
K(\Lambda_1,\Lambda_2;t)) = \frac {1}{(2 \pi it)^{(m+n)/2}}
\frac {1}{m!} \frac {1}{n!} det(e^{ i(\lambda_{1,i}-
\lambda_{2,j})^2 /2t}) det(e^{-i({\bar\lambda}_{1,\alpha}-
{\bar\lambda}_{2, \beta})^2 /2t}), \label{corto}
\end{equation}
which has the same symmetry properties of $\tilde{\xi} (\Lambda;t)$.
{}From Eqs.(\ref{propagator 2}), (\ref{K}) and (\ref{corto}) we obtain
\begin{eqnarray}
\int dU e^{ - \frac {i}{t} str(\Lambda_1 U \Lambda_2 U^\dagger)} =
factor \times
\Bigg(\frac {1}{it}\Bigg)^{mn} \ \Bigg(\frac {1}{it}\Bigg)^{-\frac {m(m-1)}{2}}
\ \Bigg(-\frac {1}{it}\Bigg)^{-\frac {n(n-1)}{2}}  \nonumber \\
\times e^{- \frac {i}{2t} str(\Lambda_1^2 + \Lambda_2^2)} \frac
{ det(e^{i (\lambda_{1,i}- \lambda_{2,j})^2 /2t}) det(e^{(-i{\bar\lambda}_{1,
\alpha}-{\bar\lambda}_{2, \beta})^2 /2t}) }{ \tilde{\Delta} (\lambda_1)
\tilde{\Delta} (\lambda_2) },
\label{integral}
\end{eqnarray}
where $factor$ is independent of $\Lambda_1$ and $\Lambda_2$.
Now, we use that $ e^{Str(ln(A))} = Sdet(A)$ with
$ A =  e^{- \frac {i}{2t} (\Lambda_1^2 + \Lambda_2^2)} $ to get
\begin{eqnarray}
\int [dU] e^{\beta Str(\Lambda_1 U \Lambda_2 U^\dagger)}  =
factor \times \beta^{mn} \left [ (\beta)^{-\frac {m(m-1)}{2}} (-\beta)^
{-\frac {n(n-1)}{2}}  \right ]  \nonumber \\
\times \frac { det(e^{\beta \lambda_{1,i} \lambda_{2,j}})
det(e^{-\beta {\bar\lambda}_{1,\alpha} {\bar\lambda}_{2,\beta}}) }
{\tilde{\Delta} (\lambda_1) \tilde{\Delta} (\lambda_2)},
\end{eqnarray}
where $\beta = \frac {1}{it}$.
$Factor$ is an arbitrary constant since it depends on $\mu$
(see eq.(\ref{measure2})). We fix it in such a way that equation (9) holds.
The implications of our formula (\ref{superint IZ 2}) for the character
expansion on $U(m|n)$, together with a detailed  calculation of the correlators
of  superunitary matrices  will be discussed  elsewhere. Here we content
ourselves with the following result
\begin{eqnarray}
\int [dU] [ Str(\Lambda U \Omega U^\dagger) ]^k = \left \{ \begin{array}{ll}
                                      0     & \mbox{ if } k=0, \ldots , mn-1 \\
                                      (mn)!\ \Sigma(\lambda,\bar\lambda)
                                      \Sigma(\omega,\bar\omega)& \mbox{ if }
k=mn.
                                                     \end{array}
                                            \right.
\label{limit}
\end{eqnarray}
The simplest identities that follow from the above  result are
\begin{eqnarray}
\int [dU] U_{i_1 j_1}U_{j_1 i_1}^\dagger...U_{i_r j_r}U_{j_r i_r}^\dagger=0,
\ \mbox{for} \ r=1,2 \ldots mn-1, \\
\int [dU] U_{11}U_{11}^\dagger...U_{mm}U_{mm}^\dagger=\frac{1}{m!},
 \ \ \mbox{for}  \ U\in U(m|1),\\
\int [dU] U_{22}U_{22}^\dagger...U_{1+n\ 1+n}U_{1+n \ 1+n}^\dagger=
\frac{(-1)^n}{n!}, \ \
\mbox{for} \ U\in U(1|n).
\end{eqnarray}

\vfill
\eject

\centerline{\bf Acknowledgements}
\noindent The authors are grateful to Professor  M.L. Mehta for a careful
reading of the manuscript. JA thanks to " Programa de Cooperaci\'on
Cient\'\i fica con Iberoam\'erica" and D. Espriu for a fruiful stay at the
Universidad de Barce\-lo\-na . He also thanks A. Gonz\'alez-Arroyo
for giving him the opportunity to visit  the Universidad Aut\'onoma de
Madrid. RM was supported by a graduate fellowship from CONICYT.
LFU  was partially supported by Fundaci\'on Andes through the Program
of Profesores Visitantes and he  is thankful for the kind hospitality of
J. Alfaro at Pontificia  Universidad Cat\'olica de Chile.  He also
acknowledges partial support from the grant UNAM-DGAPA-IN100694
and from the  grants CONACYT  400200-5-3544E and CONACYT(M\'exico)
-CONICYT(Chile) \#E120-2778.

\end{document}